\documentstyle[prd,aps,floats,epsfig,verbatim,rotating]{revtex} 
\newcommand{\be}{\begin{equation}}
\newcommand{\ee}{\end{equation}}
\newcommand{\bea}{\begin{eqnarray}}
\newcommand{\eea}{\end{eqnarray}}

\renewcommand{\Im}{\mbox{Im}}

\newcommand{\vertice}{\mbox{ 
\begin{picture}(14,6)
 \put(6,2){\circle{8}}
 \put(2,2){\line(-1,1){4}}
 \put(2,2){\line(-1,-1){4}}
 \put(10,2){\line(1,1){4}}
 \put(10,2){\line(1,-1){4}}
\end{picture}}}

\tighten 
\begin{document} 
\preprint{LPTHE-00-24}
\title{\bf THE LANDAU POLE AT FINITE TEMPERATURE} 
\author{\bf H. J. de Vega$^{(a)}$, M. Simionato$^{(a,b)}$} 
\address
{(a) LPTHE, Universit\'e Pierre et Marie Curie (Paris VI) et Denis Diderot 
(Paris VII), Tour 16, 1er. \'etage, 4, Place Jussieu, 75252 Paris cedex 05, 
France\\
(b) Istituto Nazionale di Fisica Nucleare, Rome, Italy} 
\date{\today}
\maketitle 
\begin{abstract} 
We study the Landau pole in the $\lambda\phi^4$ field
theory at non-zero and large temperatures. We show that the position
of the thermal Landau pole $\Lambda_L(T)$ 
is shifted to higher energies with respect to the zero temperature Landau 
pole $\Lambda_L(0)$. We find for high temperatures $ T>\Lambda_L(0) $,  
$\Lambda_L(T) \simeq \pi^2 \;  T/\log\left(T / \Lambda_L(0) \right) $. 
Therefore, the range of applicability in energy of the $\lambda\phi^4$ field
theory increases with the temperature.
\end{abstract} 

As is well known the $\lambda\phi^4$ field theory as well as QED suffer from
unphysical singularities at the Landau pole\cite{landau}. 
At zero temperature and resumming one-loop bubbles, this pole is found at an
energy $  \Lambda_L(0) = \mu \; e^{16\pi^2/[3\lambda] } $, where $ \mu
$ is the renormalization scale. In spite of that, 
$\lambda\phi^4$ can be used as an effective field theory for energy
scales well below $ \Lambda_L(0) $. In such a range, the theory makes
full sense and contains a rich physics\cite{zj}. The Landau pole for very small
coupling is at extremely high energies (beyond the Planck scale in the
case of QED \cite{landau}) which reduces its relevance. 
In the context of the standard model the zero temperature
Landau pole provides a limitation for the  phenomenologically
validity of the theory through the triviality bound\cite{moti}. 
In addition, in the $O(4)$ linear sigma model where $ \lambda/(4\pi) \sim 1 $
the Landau pole again limits the domain of applicability of such
effective theory.  

It is therefore an important question to assess how the Landau pole
moves when the temperature is non-zero, the Landau pole behaviour for
high temperatures $ T \gg \mu $ being especially relevant.
To the best of our knowledge this problem has not been studied in the
literature. We obtain  in this note the Landau pole position as a
function of the temperature.

The Landau pole and the phenomena linked to it  clearly correspond to
a scale of energy much higher than the renormalized mass for small
coupling. Therefore, we can  work in the massless $ \Phi^4 $ theory
without loosing generality. Let us consider the critical 
theory with euclidean bare lagrangian
\be\label{lagra}
{\cal L}= \frac{1}{2} (\partial_{\mu} \Phi_0)^2 +\frac{1}{2}
m^2_0\; \Phi_0^2(x)  + \frac{\lambda_0}{4!} \;
\Phi_0^4(x)\;, \label{lagrangian}
\ee
where $ m^2_0 $ and $ \lambda_0 $ stand for the bare mass and coupling
constant, respectively.

By massless theory, we
mean a theory where the renormalized mass, defined  as the zero
momentum two-point function, exactly vanishes
\be
m_{ren}^2(T) \equiv \Gamma_2(p^0=0,\vec p=0;T) = 0 \; ,
\ee
at a given temperature $ T $. 

Our analysis on the running of the coupling constant $\lambda(q;T)$ 
remain valid in the massive theory provided that the renormalized thermal
mass $m_{ren}(T)\neq0$ is much smaller than $T$ or that we 
consider momenta $q\gg m_{ren}(T)$. 

The renormalized mass is derived from the self-consistent gap 
equation which has the form\cite{kapusta},
\be\label{ecgap}
m^2_{ren}(T)=m_0^2 +\frac12\, \lambda_0 \, T \sum_{n=-\infty}^{+\infty}
\int\frac{d^3q}{(2\pi)^3}\frac1{\nu_n^2+q^2+
m^2_{ren}(T)},\quad \nu_n=2\pi nT\;.
\ee

We choose a negative bare mass  $ m_0^2 = - m^2 < 0 $ in order to
find a zero physical mass at some positive temperature. Setting $
m^2_{ren}(T) = 0 $ in eq.(\ref{ecgap}) yields,
$$
0 =  - m^2 + {\lambda_0 \, T^2 \over  24}
$$
Therefore, we choose from now on $ m = \sqrt{\lambda_0  \over  24}\; T
$ in order to have a vanishing physical mass. 

We want to study the evolution of the running coupling constant, defined
from the four point vertex at the four-particle static symmetric point
$\bar p_i=(0,\vec p_i)$. That is, 
\be\label{4SP}
\bar p_i^2=q^2,\quad\bar p_i\cdot \bar p_j=-q^2/3,\quad i\neq j,\quad
i,j=1,2,3,4\;,
\ee
where $ q>0 $ is a normalization mass scale. 

We have,
\be\label{coupling}
\lambda(q;T)=\Gamma_4(\bar p_1,\bar p_2,\bar p_3,\bar p_4;T)\;.
\ee
In particular we define the zero-temperature $\mu-$momentum
renormalized coupling constant as
\be\label{lambda.ren}
\lambda(\mu;0)=\lambda=
\left.\Gamma_4(\bar p_1,\bar p_2,\bar p_3,\bar p_4;0)\right.|_{q=\mu}\;.
\ee
We notice that there is a certain freedom in the choice of 
the thermal coupling constant. Other prescriptions besides
eq.(\ref{coupling}) are possible as well. 

At the one-loop level the four point function is given by the bubble
diagram \vertice plus the two crossed bubble graphs; we have the explicit
expression  at the symmetric point,
\be\label{burb}
I_{bub}(\nu_n,q;T) = \frac32\;  T\sum_{m=-\infty}^{+\infty} \int
\frac{d^3p}{(2\pi)^3} \; {1\over\nu^2_m+{\vec p\;}^2}  \; 
{1\over (\nu_m+\nu_n)^2+(\vec p+\vec q)^2} \; ,  \label{ibubble}
\ee
where a $\frac12$ comes from  the symmetry factor of the bubble
diagrams. 

We renormalize $ I_{bub} $ with a zero-temperature $\mu-$momentum
subtraction as  
\be
\Pi(\nu_n,q) = I_{bub}(\nu_n,q;T)- I_{bub}(0,\mu;0) \;.
\ee

For critical theories the renormalized bubble diagram 
$\Pi(s,q;T)$ can be analytically computed for any temperature, 
within the imaginary time formalism formalism with analytic
continuation to complex $ s =\nu_n $, or within the real time
formalism \cite{CSD} 
with the result
\be\label{bubble}
\Pi(s,q;T) = \frac{3}{(4\pi)^2}\left\{
\ln\left(\frac{\mu}{4\pi \, e \, T}\right) +\frac{s}{q} \;
\mbox{arctg}{q \over s}\; 
-\frac{i\pi T}{q} \log\left[\frac{\Gamma\left( \frac{is+q}{4\pi
iT}\right)\Gamma\left(1+\frac{is+q}{4\pi iT} \right)}{\Gamma\left(
\frac{is-q}{4\pi iT}\right)\Gamma\left(1+\frac{is-q}{4\pi iT} \right)} \right]
\right\}\; ,
\ee
where $ e = 2.7182818285\ldots $ is the basis of natural logarithms. 

For non-critical theories there are corrections 
to this result of the order $m_{ren}(T)/T$. The four point vertex
can be perturbatively evaluated as
\be
\Gamma_4(\bar p_1,\bar p_2,\bar p_3,\bar p_4; T)=\lambda-\lambda^2 \;
\Pi(0,q)+O(\lambda^3)\; . 
\ee
The one-loop resummed thermal running coupling constant 
$\lambda(q;T)$ as obtained from renormalization group analysis takes
the form
\be\label{gen.lambda}
\lambda(q;T)=\frac\lambda{1+\lambda \; \Pi(0,q;T)} \quad .
\ee
The explicit form is
\be\label{lambda.static}
\lambda(q;T)=\frac\lambda{1+\frac{3\lambda}{(4\pi)^2}\left\{
\ln\frac{\mu}{4\pi \, e \, T}+\frac{\pi T}q\left[-\pi+4\;\Im\ln
\Gamma\left(\frac{q}{4\pi i T}\right) \right] \right\}}\quad .
\ee
Notice that $ \lambda(q;T) $ is a function of $ q/T $ (and $ \mu/T
$). The low and high temperature limits follow using the Stirling formula  
and small argument expansions for the Gamma function\cite{GR}
\be
\label{lnG1}
\Im\ln\Gamma(-i x)\buildrel{x\to 0^+ }\over =\frac\pi2+\gamma \; x+{\cal
O}(x^2) \quad,
\ee
and
\be\label{lnG2}
\Im\ln\Gamma(-i x)\buildrel{x\to  \pm\infty }\over =-x\;\ln{|x|\over e}
+\frac\pi4+\frac1{12 \, x}+{\cal O}\left({1 \over
x^2}\right)\quad , 
\ee
where $ \gamma = 0.5772157\ldots $ stands for the Euler-Mascheroni constant.

We find for soft momenta $ q \ll T $,
\be
\lambda(q,T)=\frac{16}3\frac qT+O\left(\frac{q^2}{T^2}\log \frac T\mu\right)
\ee
i.e. the interaction linearly vanishes at low momenta due to the
dimensional reduction to a three-dimensional theory, as expected.

At zero-temperature we recover from eq.(\ref{lambda.static}) the usual
zero-temperature resummed coupling constant\cite{zj}
\be
\label{lambda.zero.T}
\lambda(q;0)=\frac\lambda{1+\frac3{32\pi^2}\; \lambda\ln\frac {\mu^2}{q^2}}
\ee
which exhibits the  Landau pole at the scale
\be
\Lambda_L(0) \equiv \Lambda_0=\mu \; e^{16\pi^2/[3\lambda]} \; .
\ee
In the context of effective field theories, this scale is interpreted as 
the scale where new physics enters, i.e. the scale of the order of
the ultraviolet cutoff of the theory $\Lambda$ where irrelevant operators
play an important role and therefore the theory defined by the
lagrangian (\ref{lagra}) breaks down.

A renormalization group resummed perturbation 
theory can be consistently implemented only when the
internal momenta are integrate up to a scale of order $\Lambda\leq\Lambda_0$. 

We would like to know how thermal effects affect the  Landau pole. In
particular where is the position of the thermal Landau pole  $\Lambda_L(T)$
at temperature $ T $.
This is  the principal outcome of our analysis.


We define the thermal Landau point $\Lambda_L(T)$ as the zero 
of the denominator in eq.(\ref{lambda.static}) for the resummed
coupling $ \lambda(q;T) $. That is, $\Lambda_L(T)$ is the solution of
the  equation,
\be \label{lambdaT}
{ \pi -4\;\Im\ln \Gamma\left(\frac{\Lambda_L(T)}{4\pi i T}\right)
\over \Lambda_L(T) } = {1 \over \pi \, T } \left[ \frac{(4\pi)^2}{3\lambda}
+ \ln\left(\frac{\mu}{4\pi \, e \, T}\right) \right] \; .
\ee
This is is a transcendental equation which in general
can be only solved numerically. However, in the high ($T\geq \Lambda_0$) 
and low ($T\ll \Lambda_0$) temperature limits, the position of the
Landau pole can be determined analytically using eqs.(\ref{lnG1}) and
(\ref{lnG2}), respectively.

\begin{figure}
\begin{turn}{-90}
\epsfig{file=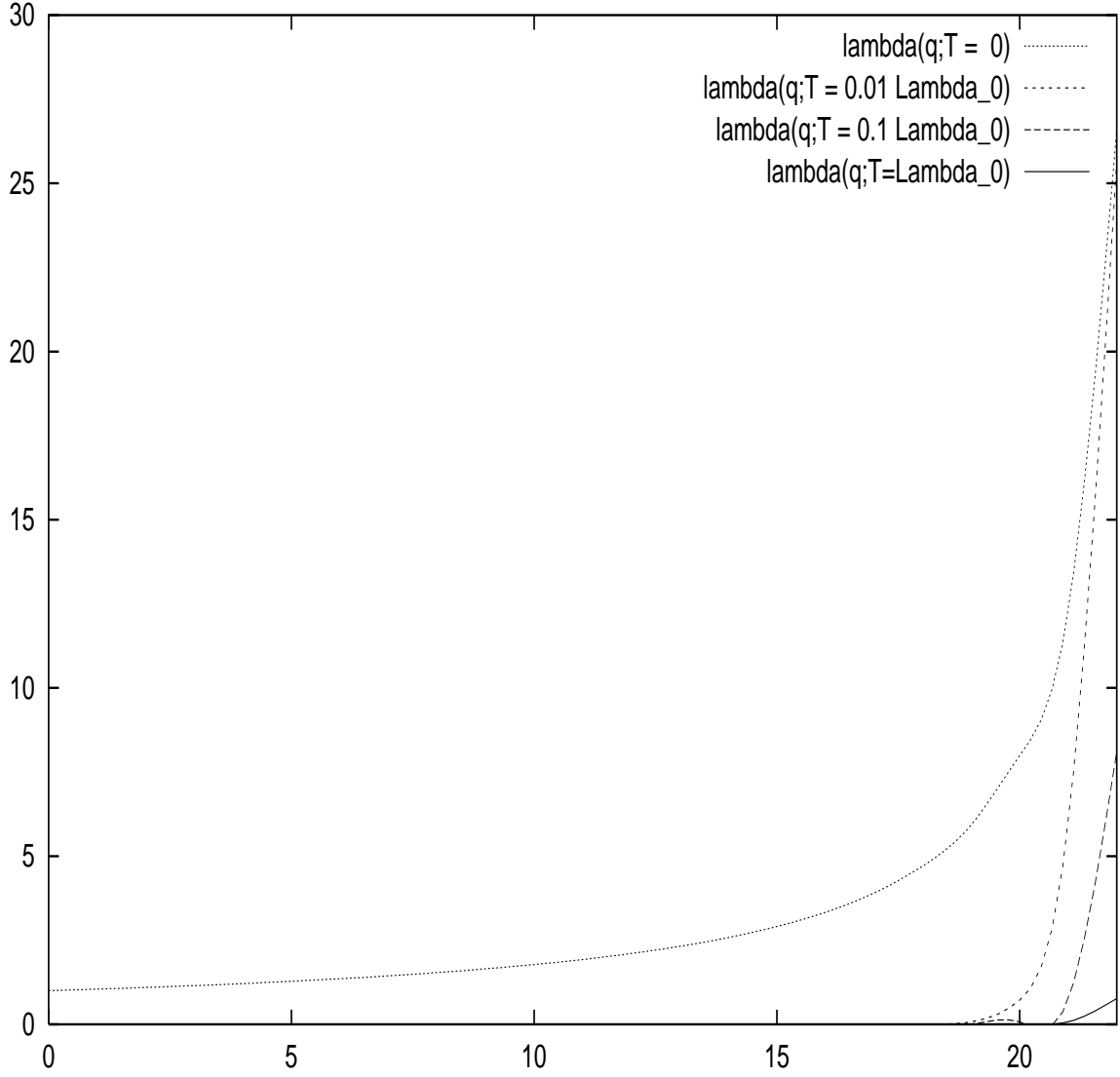,width=16cm,height=16cm} 
\end{turn}
\caption{ The resummed coupling constant  $\lambda(q;T)$ as a function
of $ x \equiv \log_{10} {q \over \mu } $ for four different
temperatures $ T = 0 $ (upmost  line), $ T = 0.01\;\Lambda_0$
(second line from above), $ T =0.1\;\Lambda_0 $ (third line from
above) and $ T =\Lambda_0 $ (bottom line). The curves correspond to  the
renormalization prescription $\lambda(\mu;0)=\lambda=1$, such that the
zero temperature Landau pole is at the scale
$\Lambda_0=7.25\cdot 10^{22} \mu$. At high
temperatures the increasing of the coupling constant at large momenta is
less pronounced that at zero-temperature and the Landau pole is shifted
to higher momentum scales. \label{fig1} }
\end{figure}

We find for temperatures below the zero temperature Landau pole
$T\ll\Lambda_0$ 
\be\label{Tchica}
\Lambda_L(T)\buildrel{T\ll\Lambda_0}\over
=\Lambda_0+\frac{4\pi^2}{3\Lambda_0}\; T^2+{\cal O}\left({T^3 \over
\Lambda_0^2}\right)\quad \; . 
\ee
That is, the temperature {\bf increases } as $ T^2/\Lambda_0 $ the
position of the Landau pole in energy above its zero temperature value.

We find for temperatures above the zero temperature Landau pole
$ T\gtrsim\Lambda_0 $, 
\be\label{Tgran}
\Lambda_L(T)\buildrel{T\gtrsim\Lambda_0}\over=\frac{\pi^2 \;
T}{\ln \left({4\pi \, e \, T \over \Lambda_0}\right)-\gamma+ 
{\cal O}\left({\Lambda_0 \over T}\right) }\quad .
\ee
We see that the thermal Landau pole position grows linearly as $
T/\ln T $ and it is therefore {\bf well above} the zero temperature
Landau pole. 

We notice that the analytic formulas (\ref{Tchica}) and  
(\ref{Tgran}) are in excellent agreement with the numerical evaluation
using eq.(\ref{lambda.static}) even beyond their expected range of
validity. In particular, eq.(\ref{Tgran}) reproduces
eq.(\ref{lambdaT}) with better than $ 1\% $ accuracy at $ T =
\Lambda_0 $ and $\lambda=1$. 

In figure 1 we display the evolution of the
resummed coupling $\lambda(q;T)$ given by eq.(\ref{lambda.static}) for
different temperatures starting from $q=\mu$ up to $q=10^{22}\mu$ for
$\lambda=1$. In figure 2 we display the position of the thermal Landau pole for
different temperatures and $\lambda=1$.

In summary, both in low $T$ and in the high $T$ region the thermal
effect consists in pushing the Landau pole scale to higher energies
than the zero temperature Landau pole. This is true for all temperatures. 

\begin{figure}
\begin{turn}{-90}
\epsfig{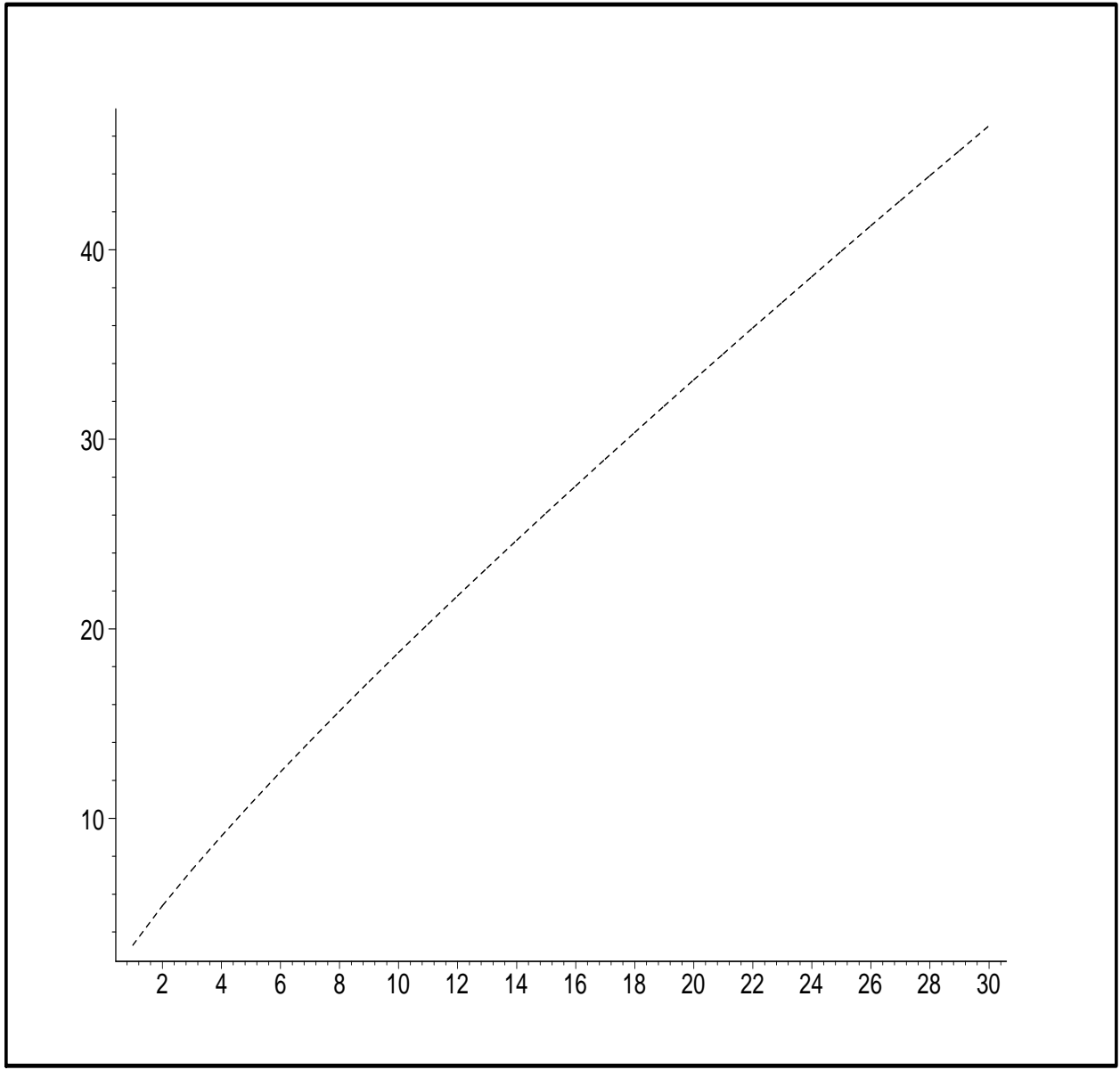} 
\end{turn}
\caption{ Position of the Landau pole $\Lambda_L(T) $ in units of 
the zero temperature pole $\Lambda_0$ as a function of temperature
from $T=\Lambda_0$ to $T=30\Lambda_0$. Here $\lambda(\mu;0)=\lambda=1$.
\label{fig2} } 
\end{figure}

The Landau pole behaves for high and low temperatures in very
different ways due to the different runnings of the coupling.
For high temperature dimensional reduction takes place yielding a
dimensionful three dimensional coupling  $ \lambda \; T $. Therefore,
the one-loop contribution to the effective coupling at momentum $ q $
behaves as  $ T/q $. This follows just by dimensional considerations or by
inspecting the one-loop diagram with two propagators integrated
over three-dimensional loop momenta. The Landau pole appears when the
one-loop contribution balances the tree level  contribution and now
this can happen when $ q \simeq T $. This gives a simple understanding
why the  Landau pole behaves as in eq.(\ref{Tgran}) for high temperature.

Since the Landau pole grows in energy when the temperature grows, the
range of validity of $ \Phi^4 $ as an effective theory does increase. 
As mentioned at the begining, the Landau pole sets the limit of
validity for theories as the standard model\cite{moti} and the $O(4)$ linear
sigma model.

As is known, fermionic systems often exhibit behaviours which differ from
those of bosonic systems (see for example \cite{kk}). We find in spinor
electrodynamics that that the Landau pole is well above $ T $ for
temperatures above the Landau pole at zero temperature. In addition,
the Landau pole position grows with $ T $ in such regime [work in
progress]. Therefore, the Landau pole behaviour in QED seems analogous to 
the scalar case.

Finally we notice that the conclusions of this paper trivially generalize to 
$O(N)-$models with interaction $ \frac{\lambda'}{2N} 
[\vec \Phi^2(x) ]^2$ through a trivial rescaling of the coupling
constant by a factor three. It is interesting to consider the large
$N$ limit since in this limit the present results on the Landau pole
are {\it exact}, even in the strong coupling regime.

\end{document}